\documentclass{article}

\usepackage{arxiv}

\usepackage[utf8]{inputenc} 
\usepackage[T1]{fontenc}    
\usepackage{hyperref}       
\usepackage{url}            
\usepackage{booktabs}       
\usepackage{amsfonts}       
\usepackage{nicefrac}       
\usepackage{microtype}      
\usepackage{graphicx}
\graphicspath{ {./images/} }
\usepackage{natbib}
\usepackage{float}
\usepackage{algorithm}
\usepackage{algpseudocode} 
\usepackage{amsmath}
\usepackage{subcaption}
\setcitestyle{numbers,sort&compress}

\title{Automated Solar Radio Burst Detection Using Deep Learning on Augmented e-Callisto Data}

\author{
\href{https://orcid.org/0009-0008-5435-2972}{\includegraphics[scale=0.06]{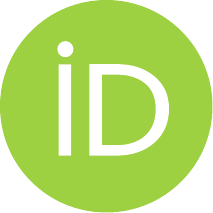}\ Vincenzo Timmel} \\
Institute for Data Science, School of Computer Science\\
University of Applied Sciences and Arts Northwestern Switzerland (FHNW)\\
5210 Windisch, Switzerland\\
\texttt{vincenzo.timmel@fhnw.ch}
\And
\href{https://orcid.org/0000-0002-5177-6875}{\includegraphics[scale=0.06]{orcid.pdf}\ André Csillaghy} \\
Institute for Data Science, School of Computer Science\\
University of Applied Sciences and Arts Northwestern Switzerland (FHNW)\\
5210 Windisch, Switzerland\\
\texttt{andre.csillaghy@fhnw.ch}
\And
\href{https://orcid.org/0000-0002-3178-363X}{\includegraphics[scale=0.06]{orcid.pdf}\ Christian Monstein} \\
Istituto ricerche solari Aldo e Cele Daccò (IRSOL)\\
Faculty of Informatics, Università della Svizzera italiana (USI)\\
6605 Locarno, Switzerland\\
}

\begin{document}
\maketitle
\begin{abstract}
\textbf{Context:}
Solar radio bursts are signatures of energetic events associated with solar flares and coronal mass ejections and can interfere with terrestrial and space-based communication systems. Real-time automatic burst monitoring enables early warnings tens of minutes to hours before associated particles reach the Earth and provides the basis for long-term statistical studies. The e-Callisto network is a worldwide system of solar radio spectrometers providing continuous 24/7 observations, with individual instruments together covering frequencies from approximately 20 MHz up to 1 GHz\footnote{The nominal per-instrument tuner range is 45--870 MHz, but individual stations are frequently upgraded with different tuners, converters, or antennas, extending coverage down to $\sim$20 MHz and up to 1 GHz at some sites; see the instrument qualification records at \url{https://www.e-callisto.org/Qualification/applidocs.html} for current per-instrument specifications.}. Detection and labeling currently rely largely on human experts, limiting scalability and real-time applicability due to hardware heterogeneity and low signal-to-noise ratios.

\textbf{Aim:}
We aim to develop an automated solar radio burst detection system that processes each newly uploaded 15-minute observation and provides near-real-time information on solar activity approximately 30 seconds after the file becomes available.

\textbf{Methods:}
We present \emph{FlareSense}, an automated detection system based on a deep Residual Network (ResNet) trained on a large manually labeled dataset. To improve robustness to out-of-distribution data and newly added instruments, we apply data augmentation techniques adapted from speech processing, including SpecAugment and TimeWarp.

\textbf{Results:}
The model achieves 93\% precision and 73.15\% recall on a manually curated test set. At the same precision as the routine expert catalog, FlareSense achieves higher recall than routine cataloging. Gradient-based explainable AI methods indicate that the network attributes its decisions to physically meaningful burst features even in the presence of noise and instrumental artifacts.

\textbf{Conclusions:}
These results demonstrate that FlareSense can automate solar radio burst detection across the entire e-Callisto network for near-real-time space-weather applications. The labeled dataset, code, and a live demo are publicly available under an MIT license.
\end{abstract}


\section{Introduction}
\label{sec:introduction}

\begin{figure}[H]
    \centering
    \includegraphics[width=0.9\linewidth]{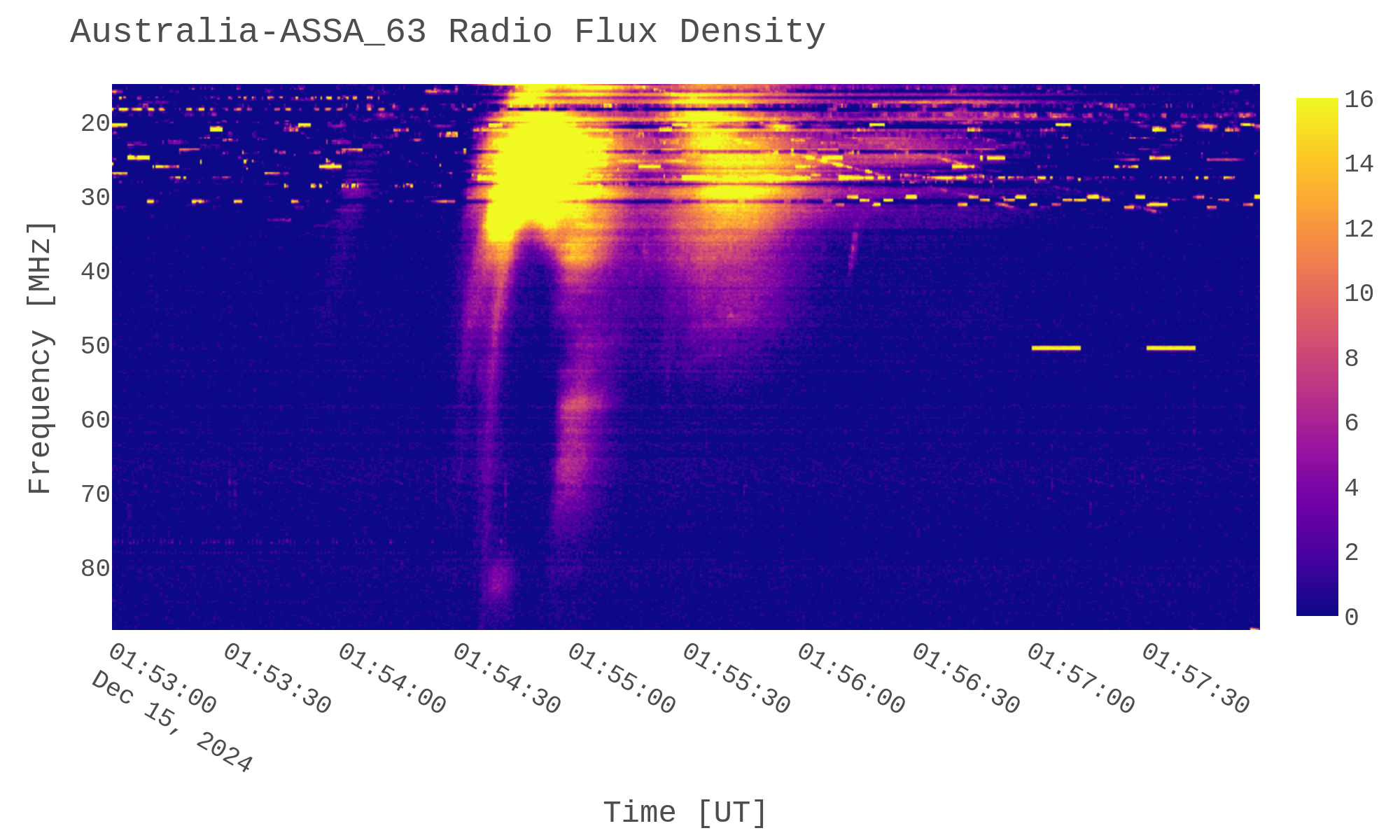}
    \caption{A solar radio burst seen by one of the instruments of the e-Callisto network in Australia.}
    \label{fig:assa_u_burst}
\end{figure}

Solar radio bursts (SRBs) are intense radio signals triggered by solar activity such as flares and coronal mass ejections, typically produced by accelerated electrons interacting with the coronal plasma. These bursts are visualized in the form of spectrograms.
A spectrogram represents the dynamic evolution of solar radio emission over time and frequency in the form of an image. The time is represented on the x-axis and the frequency on the y-axis. The pixel values correspond to the intensity of the radio emission.
Figure \ref{fig:assa_u_burst} shows such a spectrogram, including a radio burst observed by an e-Callisto station in Australia \citep{ASSA_CALLISTO}.
Characteristic morphological patterns, such as drifting structures, appear clearly and are directly related to electron beam velocities and ambient plasma density gradients. These features can serve as critical diagnostics for probing coronal magnetic fields, plasma densities, and the locations of particle acceleration sites.

Solar radio bursts can be classified into five types \citep{white2024solar}. Type I bursts are associated with noise storms from active regions. Type II bursts are linked with shock waves from coronal mass ejections. Type III bursts result from fast electron beams traveling along open magnetic field lines. Type IV bursts are thought to be related to continuum emission following flares. Finally, type V bursts may be a diffuse continuation of type III activity. 

While providing important physical observables, solar radio bursts do also significantly interfere with terrestrial and space-based radio communication systems. This is a strong motivation for continuously monitoring the Sun in the radio domain, and for the development of automated detection systems to track such events. In this context, this work does not attempt burst-type classification, but focuses solely on the setting of binary detection (burst vs. non-burst), with the aim of continuous (24/7) operation. In the deployed workflow, a new 15-minute observation is processed approximately 30 seconds after it becomes available.

This setting is provided by the e-Callisto network \citep{benz2009world}, which is a global network of radio stations designed to observe solar radio activity continuously. As of 12 April 2026, 269 CALLISTO instruments have been deployed across all continents; 113 of these have contributed data to the archive at least once, and 86 upload data regularly\footnote{Current deployment and operation counts are maintained in the official e-Callisto instrument-status document: \url{https://www.e-callisto.org/Callisto_DataStatus_Vwww.pdf}.}. The instruments in regular operation are sufficient to provide near real-time, 24/7 global coverage. The e-Callisto network monitors the Sun in the radio frequencies between 45 MHz and 870 MHz or any other range using a frequency converter. The data products generated by the e-Callisto network are spectrograms as shown in Figure \ref{fig:assa_u_burst} above \citep{Monstein_Csillaghy_Benz_2023}.

Detecting reliably the radio bursts in these spectrograms is challenging. The diversity of stations leads, both to strongly varying signal-to-noise ratio and to a diverse ‘zoo’ of radio frequency interference (RFI). Nonetheless, a comprehensive list of detected bursts, compiled manually, has been built up over several years and over all operating stations\footnote{burst catalog, by C. Monstein and partly by deARCE University of Alcalá. Available at: \url{https://soleil.i4ds.ch/solarradio/data/BurstLists/2010-yyyy_Monstein/}}. This is a valuable asset for training machine learning models.

The model we present is based on a ResNet detector \citep{he2015deepresiduallearningimage}, which we call \textit{FlareSense}. The model is trained on the e-Callisto spectrograms that are associated with the list of detected bursts. Therefore, each entry of this list is considered as a label for the associated spectrogram. The trained model is then integrated into an automated software system 
that enables near-real-time operation, and supports ongoing space weather monitoring efforts.

\section{Related Work}

Over the past few years, several methods have been proposed for the automatic detection of solar radio bursts in spectrogram data \citep{Afandi2020BurstFinderBR,9814298,48641bfb07ef4c24b30cafb9a6a59eff,AFANDI20246104,Yu_2025,Hettiarachchi2024TheAO}. Early works primarily focused on algorithmic approaches that relied on classical signal processing techniques. For instance, methods based on thresholding and mathematical morphology were developed to exploit the characteristic drifting patterns (e.g. the fast drift of type III bursts and the slower drift of type II bursts) in the time-frequency domain \citep{Jones2014, Salmane_2018}. One such method, based on the Radon transform, was developed by Lobzin et al. \citeyearpar{Lobzin2009}. All these techniques were able to detect bursts by transforming their curved trajectories into simpler geometric features (e.g. straight lines) that could be identified using standard image processing tools. However, they required careful manual tuning of parameters and were generally limited in their ability to generalize across different instruments or noise conditions.

Following these, machine learning (ML) methods were introduced to improve detection accuracy by automatically learning discriminative features from the data. ML classifiers such as support vector machines and random forests \citep{Breiman2001} have been applied to burst detection tasks \citep{Carley2020}; they provided improved robustness over purely algorithmic methods, but still depended heavily on features derived from statistical methods and were constrained by relatively small training datasets.

More recently, deep learning (DL) approaches have gained prominence. Notably, Gordo et al. \citeyearpar{BussonsGordo2023AutomaticBD} introduced the deARCE method, a deep neural network-based approach that uses a Convolutional Neural Network for both feature extraction and classification of solar radio bursts; in single-observatory mode, deARCE achieves a recall ( $\frac{TP}{TP + FN}$, where $TP$ and $FN$ denote true positives and false negatives) of around 86 \% and a precision ($\frac{TP}{TP + FP}$, with $FP$ representing false positives) of approximately 92 \%. These values correspond to a false-negative rate of 14\% and a false-discovery rate of 8\%, respectively, on a full year of data; the false-discovery rate should not be confused with the false-positive rate $\frac{FP}{FP+TN}$. The hybrid deARCE models (trained across multiple stations) further improve these metrics. The deARCE associated data are publicly available via the Astrodoncel Data Centre. \footnote{\url{https://astrodoncel.uah.es/dashboard/burst.php}}.

Independently, He et al. \citeyearpar{he2023_solar_mobilevit} showed that a transformer-based, lightweight MobileViT-SSDLite detector, combining transformer-based MobileViT blocks with an SSDLite object detection head, can detect and classify type II-V radio bursts in e-Callisto dynamic spectra with $\mathrm{AP}_{50}\approx0.78$ (i.e., the area under the precision-recall curve (obtained by varying the detection confidence threshold) at an intersection-over-union threshold of 0.5) and a recall of 92 \%, achieving competitive detection performance on selected datasets. However, both models were trained on a limited subset of instruments and with relatively small datasets, leaving open the question of their generalization to a broader range of observational conditions. 

Beyond these two representative DL detectors, several related DL/ML systems have been proposed for SRB recognition and forecasting. For example, González Orué et al. \citeyearpar{Oru2023AutomaticSR} trained a convolutional neural network on curated spectrogram data (from the CALLISTO network and a ground-based station) to automatically identify and classify solar radio bursts under low signal-to-noise and limited-label regimes. Liu et al. \citeyearpar{Liu2025DeepAL} addressed the labeling bottleneck more explicitly by combining deep active learning with a progressive convolutional GAN; they report very strong performance (e.g., 99.44\% accuracy using only 20.5\% labeled data, and per-class true-positive/false-positive trade-offs reported for multiple burst types), illustrating that query-efficient labeling can substantially reduce annotation cost while maintaining high classification quality. In a complementary direction, Wang et al. \citeyearpar{Wang2024SolarRB} focus on \emph{prediction} rather than spectrogram detection, using a multimodal model based on SOHO/MDI magnetic maps and sunspot parameters to forecast daily SRB occurrence; over 5449 days they report an accuracy of $0.898\pm 0.011$ with precision 0.923 and recall 0.934, and highlight sunspot-count thresholds as a strong predictive signal.

Finally, transfer-learning approaches have recently shown that strong performance can be obtained even with comparatively small labeled SRB sets: Roux et al. \citeyearpar{Roux2025TypeIA} fine-tune several pre-trained vision models (including YOLOv8) for Type~II/III recognition and report test-set F1 scores in the $\sim$87--92\% range, while also releasing code and processed data to support reproducibility.

Several very recent works apply modern detection architectures directly to e-Callisto/CALLISTO spectrograms. Tassan-Din et al. \citeyearpar{TassanDin2026YOLOv5} detect and classify bursts across 49 CALLISTO stations with YOLOv5 and ensemble methods (F1 = 73.8\%); they publicly release their labeled training data on Zenodo and build on the open-source YOLOv5 codebase, though we found no separate release of their own ensemble-method code. Deng et al. \citeyearpar{Deng2024RealTime} report a real-time YOLOv8-based detector (82.4\% accuracy at 140.9 fps), also with no public code or data release found. Wang et al. propose two further e-Callisto detectors, one task-aligned (TOOD, i.e., a Task-aligned One-stage Object Detection model) \citeyearpar{Wang2025TOOD} and one based on Deformable DETR \citeyearpar{Wang2025DeformableDETR}; both publicly release code and data. For Type III bursts specifically, Scully et al. \citeyearpar{Scully2023CongruentDL} combine GAN-simulated bursts with LOFAR observations to train a YOLOv2 detector (mAP = 77.71\%), with no code or data release found.

A further shortcoming in the literature is that, despite these advances, the detection of solar radio bursts, particularly in the context of networks such as e-Callisto, is still largely performed manually by experts. In addition, while some operational systems exist (for example, one deployed by the HUMAIN group \footnote{\url{https://www.sidc.be/humain/callisto_burst_archives_AI}} and other systems developed in China or the one by Gordo et al., see further above), an SRB detector whose code, dataset and model are all open-sourced together remains elusive.

Our work builds on these developments by leveraging deep learning and an extensive collection of e-Callisto spectrograms from multiple e-Callisto instruments to develop a model that generalizes across diverse observational conditions. Because e-Callisto data are often affected by noise, calibration differences and a diverse ‘zoo’ of radio frequency interference, we further investigated the use of data augmentation methods to enhance model robustness. Specifically, we incorporated \textit{SpecAugment} and \textit{TimeWarp} \citep{Park_2019}, adapting these techniques for solar radio spectrograms to improve the model’s ability to handle real-world variability in RFI and changing / updating instruments.

\section{SpecAugment and TimeWarp for Spectrogram Augmentation}
\label{sec:augmentation_methods}

\begin{figure}[H]
\centering
\includegraphics[width=1\columnwidth]{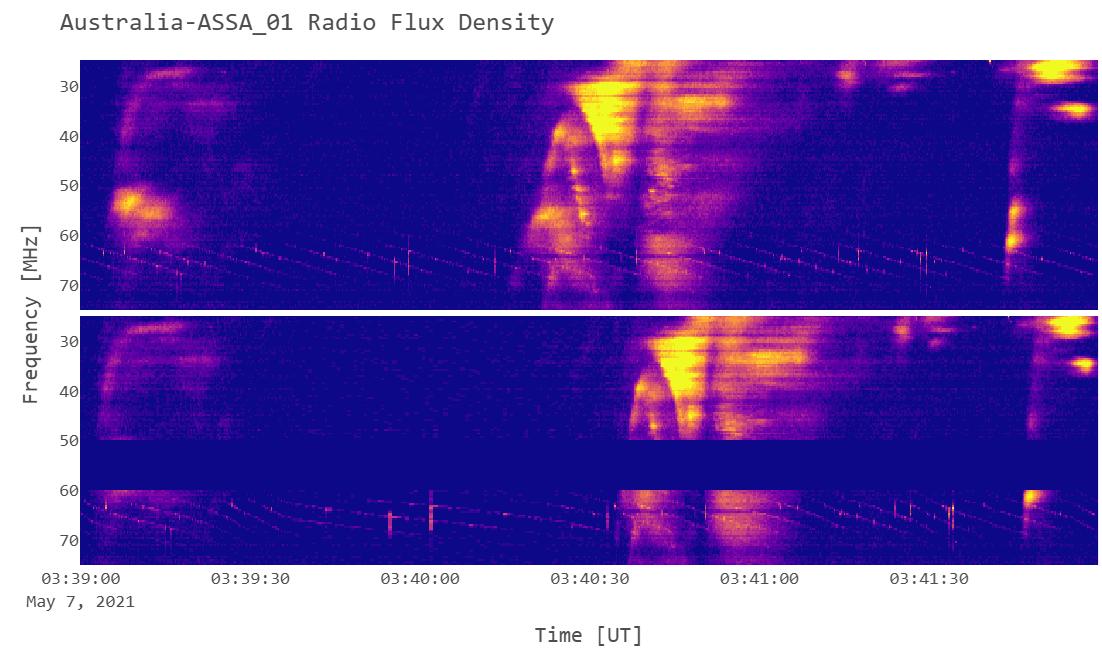}
\caption{\small 
Examples of spectrogram data augmentation methods being applied to a e-Callisto spectrogram.
Top: Unaugmented spectrogram, showing a solar radio burst seen by the instrument \textit{AUSTRALIA-ASSA\_01} on the 7th of May 2021. This example is shown for illustration only and is not part of the dataset described in Section~\ref{subsec:selected_stations}.
Bottom: \textit{TimeWarp} smoothly deforms the spectrogram along the time axis, imitating variations in burst duration and drift rate and \textit{SpecAugment} masks a random frequency band with 0, simulating missing or distorted information and improving robustness to frequency-dependent interference. }
\label{fig:augmentation_examples}
\end{figure}

Originally introduced in the field of speech recognition, \textit{SpecAugment} modifies the time-frequency representation of a signal by uniformly masking random bands along the time or frequency axes. The masked regions can be filled with a constant value, the mean/max/min amplitude of the spectrogram, or random values (typically sampled uniformly between 0 and 1), depending on which strategy yields the best performance for the task.
This simulates missing or distorted information and encourages the model to learn features that remain robust under such perturbations. 
\textit{TimeWarp}, in contrast, smoothly stretches or compresses regions of the spectrogram along the time axis, imitating variations in event duration or drift rate. 
Together, these augmentation techniques enhance the diversity of the training data and help the model generalize to heterogeneous and noisy observations, such as across different instruments or phases of the solar cycle. Figure~\ref{fig:augmentation_examples} illustrates examples of these transformations.

Although SpecAugment and TimeWarp are commonly applied to spectrograms, this is the first use of these methods on spectrograms generated from voltage signals induced by solar radio emission in antenna systems such as e-Callisto.

\section{Data}

\subsection{Spectrograms}

The raw spectrograms used in this work are obtained from stations participating to the e-Callisto network. They are archived at FHNW, which maintains the global e-Callisto public archive \footnote{The raw spectrograms of e-Callisto can be accessed at \url{https://soleil.i4ds.ch/solarradio}}. Each spectrogram represents a 15-minute observation that captures the dynamic evolution of solar radio emission over time and frequency. The time axis is sampled mostly at a resolution of 250 ms. In contrast, the frequency axis generally spans a broad range, from approximately 20 MHz up to 1 GHz, although the exact frequency range and resolution may vary from one instrument to another. For instance, while many stations operate with a spectral resolution on the order of tens of kilohertz (commonly around 62.5 kHz per channel), differences in hardware configuration can lead to variations in the number of frequency channels available in each spectrogram.

\subsection{Binary Labels}
\label{subsec:labels}
The fact that SRBs have been labeled by hand over many years is of extreme utility for training a model. The burst catalog introduced already in Section \ref{sec:introduction} can be considered as the ground truth. For our method the SRB classes are ignored, and spectrograms containing a burst are given the binary label \textit{1} and spectrograms without a solar radio burst are given the binary label \textit{0}. In this work, we restrict the task to binary classification (burst vs. non-burst) but extending FlareSense to type-specific classification is planned as future work.

Nevertheless, as with any manual labeling process, ambiguous cases may occur. Therefore, the labels are not perfectly consistent and this should be considered during data creation, training, and evaluation.

\subsection{Selected Stations and Instruments}
\label{subsec:selected_stations}

The ground truth burst catalog mentioned above contains the station which has observed the solar radio burst. However, more than one instrument might be observing at the same station, and several instruments might perform the same observation. To uniquely associate the label in the burst catalog with a single spectrogram, we select the best instruments that have performed the specific observation, as seen in Table \ref{table:bursts}.

\begin{table}[h]
\caption{Detected Radio Bursts (excerpt from the e-Callisto burst catalog). The \textit{Type} column follows the standard classification (I--V); \texttt{RBR} denotes a fixed-frequency radio burst. The full list of event codes used in the catalog is documented at \url{https://e-callisto.org/eventSummaryExplanation.pdf}. The station name is given (e.g., \texttt{Australia-ASSA}), but the exact instrument identifier, typically appended as a number (e.g., \texttt{Australia-ASSA\_01}), is not specified in this list.}
\label{table:bursts}
\centering
\begin{tabular}{llll}
\hline\hline
Date & Time (UT) & Type & Stations \\
\hline
2024-03-03 & 03:20--03:20 & III & Australia-ASSA, INDIA-UDAIPUR \\
2024-03-03 & 14:28--14:28 & RBR & Germany-DLR \\
\hline
\end{tabular}
\end{table}

To construct the dataset, we examined how often each instrument appears in the burst catalog and selected 26 instruments such that no two are from the same station recording at the same time. This ensures continuous global coverage while avoiding redundant data (and perhaps data leakage) from co-located instruments. The selected instruments, their stations, and their geolocations are listed in Table \ref{appendix:stations_instruments_combined} (Appendix).

Detailed specifications of the stations and instruments used, including sensitivity, resolution, and frequency coverage, are available on the e-Callisto page\footnote{See \url{https://www.e-callisto.org/Qualification/applidocs.html} and \url{https://www.e-callisto.org/GeneralDocuments/Callisto-General.html} for more information.}.

\subsection{Generation of Training Data}
\label{subsec:creation_training_data}

Using the burst catalog described in Subsection~\ref{subsec:labels}, which provides the burst time and instrument, we extract the corresponding raw data and construct spectrograms containing each event. To reduce positional bias, the start is randomly shifted by a uniform offset in $[0,10]$ minutes before the annotated time. The full procedure is given in Algorithm~\ref{alg:generate_burst_spectro} (Appendix).

For each burst spectrogram produced from the catalog, we additionally create ten spectrograms without a burst to provide negative examples and maintain a realistic class balance. We sample random 15-minute windows and discard any window that overlaps a burst reported in the catalog by \emph{any} station, not only by the instrument being sampled; this is deliberately conservative, since a burst seen elsewhere in the network may well be faintly present in the window at hand. The remaining windows are kept if they meet basic data-quality criteria; see Algorithm \ref{alg:generate_non_burst_spectro} (Appendix).

Overall, this method with the selected instruments and using the burst catalog up to May 2024 yields \textit{304,750} total spectrograms. The dataset is publicly available on Hugging Face\footnote{e-Callisto Radio Sunburst Dataset. Available at: \url{https://huggingface.co/datasets/i4ds/ecallisto_radio_sunburst}.}.

For all data processing, the Python package \textit{ecallisto\_ng} \citep{vincenzo_timmel_2024_14505852} was used.

To create a train, validation and test set, we perform a stratified split into 80\% training, 10\% validation, and 10\% test (stratification was done by instrument and by class label).

Although spectrograms are sampled randomly across multiple years (2022-2024) and instruments, the dataset does not explicitly control for the phase of the solar cycle. This limitation arises because no consistently labeled data exist across multiple solar cycles. Each 15-minute spectrogram is normalized independently; this largely mitigates long-term variations in background solar flux.

Still, assessing how solar-cycle-dependent variations in burst rates affect model generalization remains an open question.

Additionally, the test set labels were re-verified by the e-Callisto principal investigator (PI) in a dedicated second pass, in which every test spectrogram was inspected again without time pressure. The PI received the test spectrograms for re-inspection without access to the model predictions, so this second pass was blind to the model output. As a result, the test set contains fewer mislabeled (noisy) samples than the training and validation sets. We refer to these re-verified labels as the \textit{clean} test labels, and use them as ground truth throughout Section~\ref{sec:evaluation}.

Before passing a training spectrogram to the model, it is preprocessed as shown in Algorithm \ref{alg:data_augmentation}. Validation and test spectrograms undergo the same median subtraction, resizing, and normalization, but not the stochastic \textit{TimeWarp} or \textit{SpecAugment} steps.

\begin{algorithm}[H]
\caption{Spectrogram Augmentation Pipeline}
\label{alg:data_augmentation}
\begin{algorithmic}[1]
\Require Training dataset $\mathcal{D}=\{S_i\}_{i=1}^N$ of raw spectrograms
\Ensure Augmented dataset $\mathcal{D}_{\text{aug}}=\{S^{(\text{aug})}_i\}_{i=1}^N$, each of shape $128 \times 512$
\For{each spectrogram $S$ in $\mathcal{D}$}
    \State $S \gets \textsc{MedianSubtract}(S)$ \Comment{subtract median per frequency channel}
    \State $S \gets \textsc{TimeWarp}(S;\ w)$ \Comment{elastic deformation along time axis}
    \State $S \gets \textsc{Resize}(S,\ 128,\ 512)$ \Comment{\texttt{torchvision.transforms.Resize}}
    \State $S \gets \textsc{SpecAugment}(S;\ m_f, m_t)$ \Comment{mask random frequency/time bands}
    \State $S \gets \textsc{NormalizeToUnitInterval}(S)$ \Comment{scale pixel values to $[0,1]$}
    \State $S^{(\text{aug})} \gets S$
    \State append $S^{(\text{aug})}$ to $\mathcal{D}_{\text{aug}}$
\EndFor
\end{algorithmic}
\end{algorithm}

\section{Model for FlareSense}
A ResNet model \citep{he2015deepresiduallearningimage} was adapted for binary classification. The network was initialized with random weights using standard Kaiming normal initialization, which He et al. \citeyearpar{He2015} demonstrated enables deep feedforward ReLU networks-including ResNet-to converge faster and more stably toward a low loss. 

Weighted binary cross-entropy ($\mathrm{BCE}_{\text{weighted}}$) was used as the loss function, defined as  
$\mathrm{BCE}_{\text{weighted}} = -[\, w_1\, y \log(\hat{y}) + (1 - y)\log(1 - \hat{y}) \,]$,  
where $y$ and $\hat{y}$ denote the true and predicted labels, respectively, and $w_1$ is the weight applied to the positive (burst) class.  
The weight was computed from the ratio of burst to non-burst samples, yielding $w_1 \approx 10$ because we defined that for each burst sample, there should be 10 non-burst samples, see Subsection \ref{subsec:creation_training_data}.

A low $\mathrm{BCE}_{\text{weighted}}$ indicates that the model assigns large positive logits (and thus high probabilities) when a spectrogram contains a solar radio burst, and large negative logits when it does not.

The optimal ResNet architecture and training settings were determined through hyperparameter tuning, i.e., systematic optimization of parameters such as learning rate and weight decay, using the validation set (see Section~\ref{subsec:hyperparameter_sweep}). The sweep trained candidate models on the training split and evaluated them on the validation split. The test set was not used for model selection or hyperparameter tuning. The single temperature scalar used for probability calibration was fitted separately on the training-set predictions (Appendix~\ref{app:prob_calib}).

\subsection{Training Method}

The model was trained for a fixed number of epochs using the AdamW optimizer \citep{loshchilov2017decoupled}, which decouples weight decay from the gradient update, leading to more stable convergence compared to the standard Adam optimizer. 

A linear learning rate decay with an initial warm-up phase was applied. The warm-up prevents instability during early training by starting with a smaller learning rate and gradually increasing it before decaying linearly. This scheduling strategy is commonly adopted in transformer-based models and contributes to smoother optimization and better generalization performance in neural networks \citep{you2019how}.

\subsection{Parameters and Hyperparameter Sweep}
\label{subsec:hyperparameter_sweep}
A Bayesian hyperparameter tuning \citep{snoek2012practicalbayesianoptimizationmachine} (using \textit{Weights \& Biases} \citep{wandb}) maximized the F1 score on the validation set. The F1 score, defined as the harmonic mean of precision and recall, was computed as an unweighted mean across all instruments; meaning that each instrument contributes equally, regardless of how many bursts it has detected. This choice was made to prioritize model robustness across diverse instruments.

Table \ref{tab:params_sweep} details the parameters used for \textit{FlareSense} and the best-performing configuration identified during the sweep, i.e. the configuration with the highest validation F1 score. This configuration reached an F1 score of 80.53 \% on the validation set.

\begin{table}[h]
\caption{Model Parameters, Ranges, and Best Run Values}
\label{tab:params_sweep}
\centering
\begin{tabular}{lll}
\hline\hline
\textbf{Parameter} & \textbf{Range/Value} & \textbf{Best Run Value} \\
\hline
Epochs & 15--100 & 25 \\
Warm-up Epochs & 3--20 & 12 \\
Learning Rate  & \(1\times10^{-6}\)--\(1\times10^{-3}\) & \(2.4\times10^{-4}\) \\
Label Smoothing & 0.0--0.2 & 0.1174 \\
Model Type & \texttt{[ResNet34, ResNet50, ResNet101, ResNet152]} & \texttt{ResNet34} \\
Weight Decay & \(1\times10^{-9}\)--\(1\times10^{-3}\) & \(5.2\times10^{-4}\) \\
Frequency Masking Parameter & 0--40 & 25 \\
Time Masking Parameter & 0--90 & 70 \\
TimeWarp Width & 300--450 & 389 \\
Masking Method & \texttt{[Mean, Random, Max, Min]} & \texttt{Random} \\
Batch Size & 64 & 64 \\
\hline
\end{tabular}
\end{table}

\subsection{Effectiveness of Data Augmentation}
\label{subsec:effectiveness_augmentation}

Feature-correlation and ablation analyses from the parameter sweep indicated that data augmentation had a strong positive influence on model performance. In particular, the Bayesian optimization sweep, which adaptively explores hyperparameter combinations by balancing exploration and exploitation, consistently selected configurations with data augmentation enabled.

\begin{figure}[H]
\centering
\includegraphics[width=0.7\columnwidth]{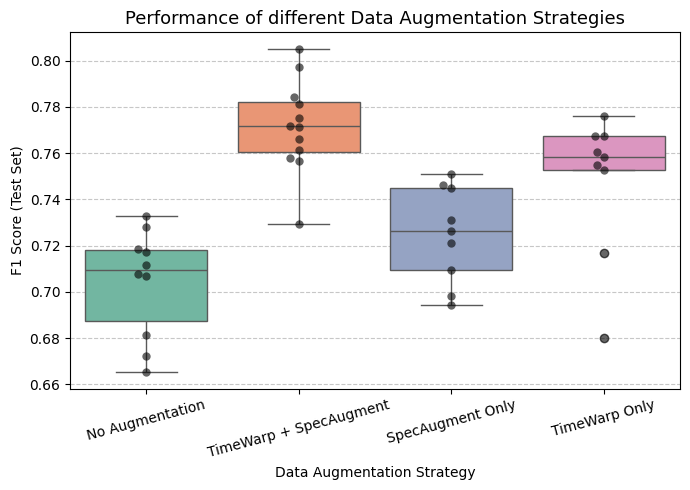}
\caption{\small 
Comparison of FlareSense test-set performance under different data augmentation strategies. Each box represents the distribution of test F1-scores across models produced by hyperparameter sweeps with identical search ranges (see Section~\ref{subsec:hyperparameter_sweep}); model optimization within each sweep used the validation split. The results show that both \textit{SpecAugment} and \textit{TimeWarp} independently improve detection accuracy, while their combination yields the highest median performance and lowest variance across experiments.}
\label{fig:boxplot_data_augm}
\end{figure}

To further quantify this effect, three additional hyperparameter tuning runs were performed using the same parameter ranges described in Section~\ref{subsec:hyperparameter_sweep}: one with data augmentation disabled, one with only \textit{SpecAugment}, and one with only \textit{TimeWarp}. These were compared to the reference run from Section~\ref{subsec:hyperparameter_sweep}, which allowed both augmentations simultaneously. Candidate models were trained on the training split and compared during the sweeps on the validation split; Figure~\ref{fig:boxplot_data_augm} reports their subsequent performance on the held-out test data. The median (and maximum) test F1-scores for the four conditions were 71.0 \% (73.3 \%), 72.6 \% (75.1 \%), 75.8 \% (77.6 \%), and 77.2 \% (80.5 \%) for no augmentation, \textit{SpecAugment} only, \textit{TimeWarp} only, and both augmentations, respectively.

This confirms that both augmentation techniques individually improve performance, while their combination yields the highest accuracy and robustness across runs.

\subsubsection{Effectiveness on Unseen Instruments}
\begin{figure}[H]
\centering
\begin{subfigure}{0.47\columnwidth}
  \includegraphics[width=\linewidth]{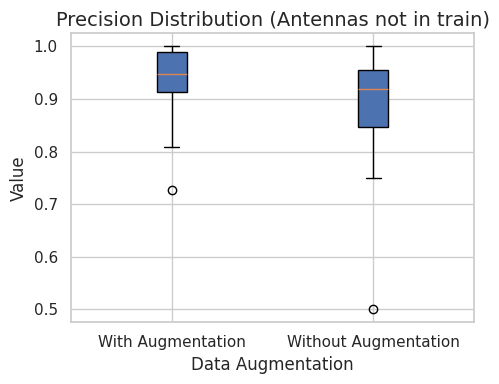}
\end{subfigure}
\hfill
\begin{subfigure}{0.47\columnwidth}
  \includegraphics[width=\linewidth]{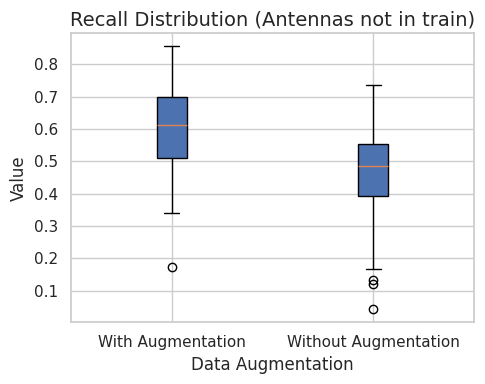}
\end{subfigure}
\caption{\small Left: Validation precision across the 17 instruments withheld from training for models trained with and without data augmentation.
Right: Corresponding validation recall for the same comparison.
Both panels show results only for instruments unseen during training, illustrating how data augmentation improves cross-instrument generalization.}
\label{fig:prec_recall_unseen_instruments}
\end{figure}

Additionally, we evaluated how data augmentation influences predictive quality on instruments excluded from model training (Figure \ref{fig:prec_recall_unseen_instruments}). For this separate diagnostic experiment, nine instruments were randomly retained in the training subset: \texttt{GLASGOW\_01}, \texttt{BIR\_01}, \texttt{ALASKA-HAARP\_62}, \texttt{ALASKA-COHOE\_63}, \texttt{AUSTRIA-UNIGRAZ\_01}, \texttt{Australia-ASSA\_62}, \texttt{HUMAIN\_59}, \texttt{INDIA-OOTY\_02}, and \texttt{SWISS-Landschlacht\_62}. The other 17 instruments were withheld from training and retained in the validation data. We then compared otherwise equivalent models trained with and without augmentation on those 17 unseen instruments. This setting reflects the regular addition of new e-Callisto instruments and changes to the calibration of existing instruments.

Figure \ref{fig:prec_recall_unseen_instruments} shows that data augmentation improves validation precision and recall on the 17 instruments excluded from training and reduces the per-instrument spread of both metrics. While precision and recall improved across most instruments, a few outliers remain, which likely reflect unique calibration or interference patterns.

\subsection{Training Process}
The final model is trained with the parameters presented in Table \ref{tab:params_sweep} on an A4500 with 20 GB of VRAM. Once the hyperparameters were fixed by the sweep, the final model was retrained on the union of the training and validation splits (90\% of the data) for the selected number of epochs, so that no data is left unused; the test set is reserved for the final evaluation. It took 7 hours and 32 minutes to fully train FlareSense.

\section{Evaluation}
\label{sec:evaluation}

The dataset is severely unbalanced in both class distribution and instrument allocation; hence, we report both micro and macro metrics, stratified per instrument, for a more accurate assessment. We also compare our model's performance against the routine expert catalog to highlight potential advantages or drawbacks.

The human baseline is the burst catalog itself as it is produced in routine operation: the PI inspects the incoming spectrograms day by day and enters the bursts that are noticed at the time. These day-to-day catalog entries are the same labels used for training, and we score them against the clean test labels obtained in the dedicated second pass described in Subsection~\ref{subsec:creation_training_data}. The gap between the two therefore quantifies how many bursts are missed under realistic operational conditions, as opposed to careful re-inspection.

To compare FlareSense with routine expert cataloging, we calibrated FlareSense’s probability outputs using temperature scaling fitted on the training-set predictions; see Appendix~\ref{app:prob_calib} and Figure~\ref{fig:prob_calib}. For the matched-precision comparison on the clean test set, we then used a calibrated probability threshold of 0.426, at which FlareSense has the same precision as the routine catalog baseline.
As shown in Table \ref{tab:fs_vs_human}, FlareSense achieves a recall of 73.15\%, compared with 63\% for the routine catalog. Thus, at equal precision, FlareSense recovers more of the bursts that were missed during routine day-to-day inspection.

\begin{table}[ht]
\centering
\caption{Performance on the Clean Test Set (n=30,549). }
\label{tab:fs_vs_human}
\begin{tabular}{lcc}
\hline
\hline
\textbf{Method} & \textbf{Precision} & \textbf{Recall} \\
\hline
FlareSense (Ours) & 93\% & 73.15\% \\
Routine expert catalog  & 93\% & 63\% \\
\hline
\end{tabular}
\end{table}

\begin{figure}[H]
\centering
\begin{subfigure}{0.45\columnwidth}
  \centering
  \includegraphics[width=\linewidth]{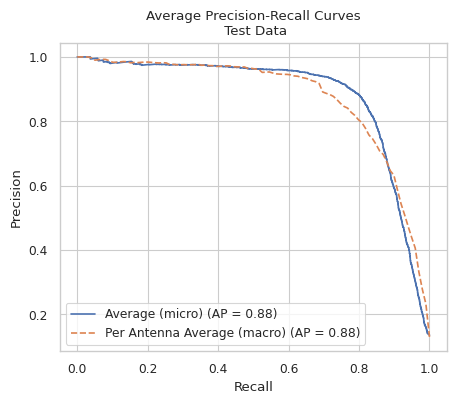}
  \caption{Test set}
  \label{fig:pres_recall_test}
\end{subfigure}
\hfill
\begin{subfigure}{0.45\columnwidth}
  \centering
  \includegraphics[width=\linewidth]{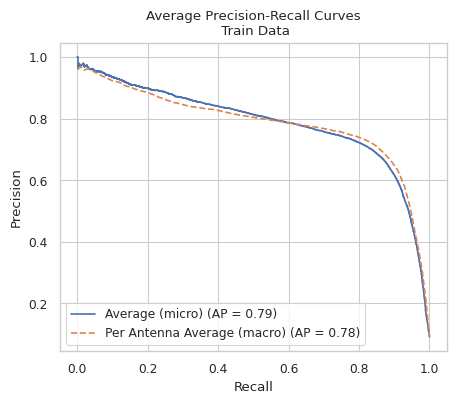}
  \caption{Train set}
  \label{fig:pres_recall_train}
\end{subfigure}
\caption{\small
Precision-recall (PR) curves for the train and test datasets.
Each point corresponds to a different probability threshold.
Macro PR values are computed by averaging precision and recall across instruments (equal weight per instrument), while micro PR values are calculated globally over all samples (weighting instruments by the number of examples).
The higher test-set curve reflects the cleaner, double-checked labels rather than a difference in generalization.}
\label{fig:pres_recall}
\end{figure}

Figure~\ref{fig:pres_recall} shows the precision-recall (PR) curves for the \textit{Test} and \textit{Train} datasets.  
The cleaner, double-checked test labels lead to a higher and more stable PR curve, showing that the model performs better when evaluated against verified labels.
In contrast, the training set’s weaker performance reflects the noisier single-checked labels used during optimization. 
We use PR rather than ROC curves because precision-recall plots are more informative for imbalanced datasets such as ours.

Generally, most of the false negatives are either very faint, Type III bursts or missed during annotation. If a burst is clearly visible and the model misses it at the selected operating point, the probabilities are often close to the relevant decision threshold. Selecting a lower threshold would also capture some of those cases if the user wishes to trade precision for recall.

\section{Model Interpretability}
\label{sec:model_robust}

\begin{figure}[H]
\centering
\includegraphics[width=1\textwidth]{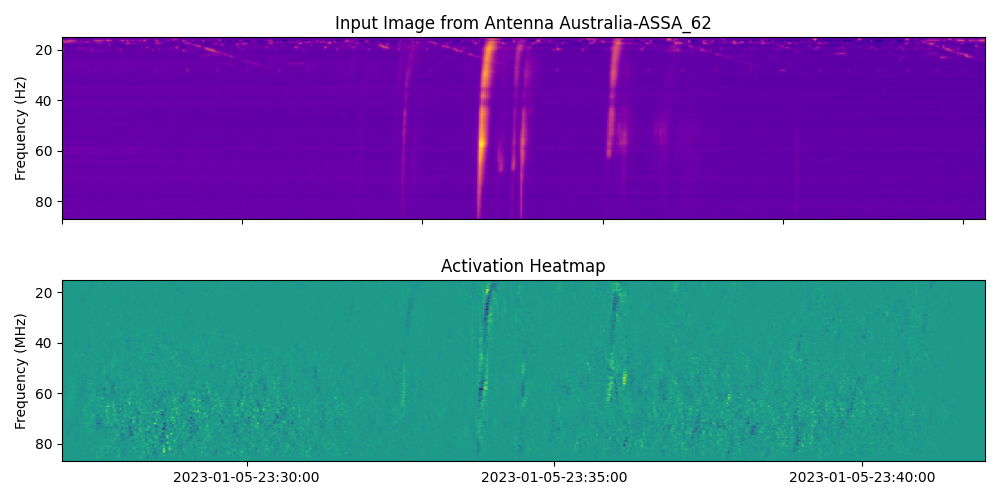}
\caption{A true positive detection by FlareSense, with the corresponding Gradient SHAP attribution map below. The attribution is concentrated on the solar radio burst structure.}
\label{fig:heatmap_8803_Australia-ASSA_62_burst}
\end{figure}

We assessed the interpretability of FlareSense using Gradient SHAP \citep{NIPS2017_8a20a862,kokhlikyan2020captum}, which attributes each model prediction to specific spectro-temporal features (i.e., pixels on the spectrogram). These maps do not by themselves establish causal model behavior, but they allow us to check whether the prediction is attributed to the burst structure or to unrelated image regions. Figure \ref{fig:heatmap_8803_Australia-ASSA_62_burst} shows a typical true positive example: the attribution map concentrates on the intense, short-duration emission band that defines the burst. This is consistent with the model using physically meaningful burst features rather than incidental structure at the edges or in the background of the spectrogram.

\begin{figure}[H]
\centering
\includegraphics[width=1\textwidth]{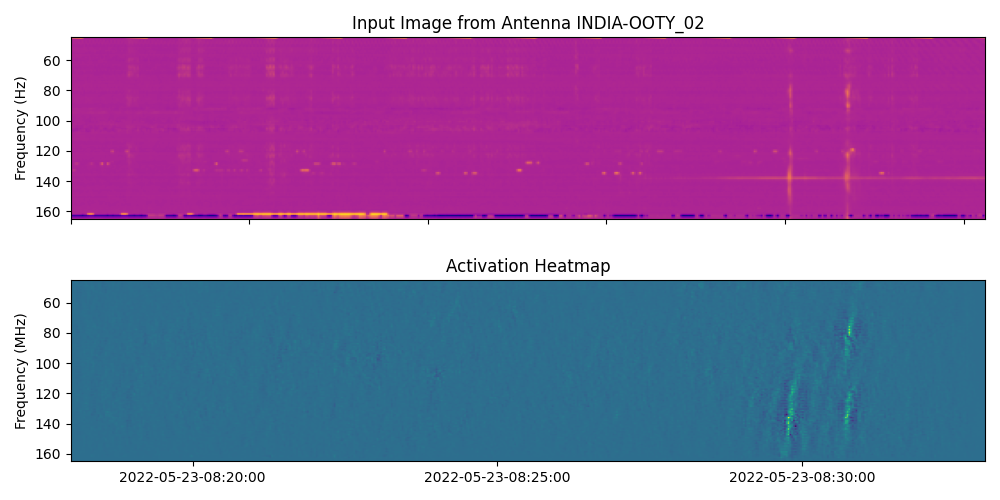}
\caption{False negative by FlareSense. Before temperature scaling, the model assigns a burst probability of 45.9\%, below the uncalibrated threshold of 0.5. With the fitted temperature, this corresponds to a calibrated probability of approximately 41.8\%, which is also below the calibrated evaluation threshold of 0.426.}
\label{fig:heatmap_20295_INDIA-OOTY_02_burst_FN}
\end{figure}

In contrast, Figure \ref{fig:heatmap_20295_INDIA-OOTY_02_burst_FN} illustrates a false negative. Although the strongest attribution falls in the relevant region, it is weak, which is consistent with the signal’s low intensity or overlapping background reducing the model confidence. However, this event might not be a true burst, as there is essentially no frequency drift. The attribution pattern nevertheless follows a drift-like direction that is not clearly visible in the instrument spectrogram. As such, this example might represent a mislabeled data point, although this is difficult to determine from a single spectrogram.

\begin{figure}[H]
\centering
\includegraphics[width=1\textwidth]{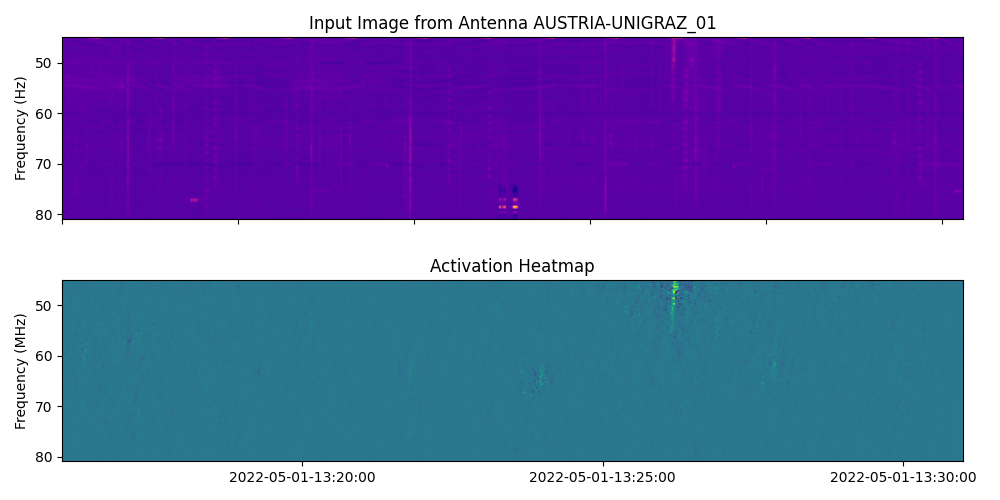}
\caption{Apparent false positive originating from a ground-truth annotation miss. FlareSense assigns a burst probability of 87.73\%, and the attribution map concentrates on the burst-like emission.}
\label{fig:heatmap_7229_AUSTRIA-UNIGRAZ_01_FP}
\end{figure}

During attribution-based review, we identified cases where the model highlighted physically plausible burst structure despite a negative label. Figure~\ref{fig:heatmap_7229_AUSTRIA-UNIGRAZ_01_FP} shows one such instance: the model score is high (87.73\%) and the attribution aligns with the emission band; upon manual reinspection, we concluded that this burst was missed during ground-truth creation and should be relabeled. We stress that no test-set label was changed as a result of these attribution-based inspections: the evaluation in Section~\ref{sec:evaluation} is computed against the clean labels produced in the PI's independent second pass (Subsection~\ref{subsec:creation_training_data}), which was carried out without access to the model predictions. The cases discussed here are therefore reported as qualitative observations, and they mean that the precision of 93\% in Table~\ref{tab:fs_vs_human} is, if anything, a conservative estimate.

Across the examples in Appendix \ref{appendix:ex_ai_examples}, the attribution maps consistently highlight visually meaningful structures. This provides qualitative evidence that the model uses burst-like signal patterns in these cases rather than unrelated regions of the images.
Per-instrument F1, precision, and recall are reported in Table \ref{tab:all_instr_metrics_percent} (Appendix~\ref{appendix:per_instrument_metrics}). In general, XAI also clarified whether a case was a genuine false positive or a missed burst.
When the model emphasized regions that appeared physically significant, further inspection often revealed that what was initially labeled as a false positive was in fact a true burst we had overlooked.

\section{Discussion \& Conclusion}
In this work, we introduced \textit{FlareSense}, a ResNet-based deep learning model designed to automatically detect solar radio bursts in e-Callisto spectrogram data. By adapting SpecAugment and TimeWarp to solar radio spectrograms and performing a hyperparameter sweep, our model matches the precision of the routine expert catalog while improving recall. Specifically, FlareSense reaches a precision of 93\% and a recall of 73.15\% on a clean, double-checked test set, compared with 63\% recall for routine cataloging at the same precision.

Our study demonstrates that incorporating data augmentation significantly enhances model robustness, especially when generalizing to unseen instruments. The ablation over augmentation strategies (Fig.~\ref{fig:boxplot_data_augm}) shows that runs without any augmentation are clearly the weakest, and that enabling \textit{SpecAugment} and \textit{TimeWarp} together yields both the highest median F1 score and the smallest spread across sweep runs. The per-instrument comparison on unseen instruments (Fig.~\ref{fig:prec_recall_unseen_instruments}) points in the same direction: augmentation raises precision and recall for most instruments and narrows their spread, underscoring the effectiveness of domain-adapted augmentation strategies.

Additionally, the explainable AI (XAI) analysis provides qualitative evidence that, in the inspected examples, predictions are attributed to physically plausible emission structures rather than unrelated image regions.

While our work addresses challenges such as class imbalance and noisy labels, it also opens several avenues for future research. Future efforts could explore more advanced techniques for handling label noise, such as semi-supervised or self-supervised learning methods \citep{berthelot2019mixmatchholisticapproachsemisupervised,sohn2020fixmatchsimplifyingsemisupervisedlearning,zbontar2021barlowtwinsselfsupervisedlearning} tailored to spectrogram data. Li et al. \citeyearpar{Li2022SelfSupervised} and Zhao et al. \citeyearpar{Zhao2025VitDetSRBs} have already applied self-supervised pretraining (masked-autoencoding and stable-diffusion-based augmentation, respectively) to solar radio spectrograms; the former publicly releases code and data, and the latter's formal data-availability statement reports no datasets generated, though the same group has released related code separately. Additionally, further refinement of augmentation methods and model architectures may yield even higher performance, especially in scenarios with extremely low signal-to-noise ratios.

Overall, \textit{FlareSense} represents a step toward fully automated solar radio burst detection, reducing the need for labor-intensive manual analysis and contributing to improved real-time space weather monitoring. \textit{FlareSense} is available at \url{https://huggingface.co/i4ds/flaresense-v2} and the code to reproduce our results at \url{https://github.com/i4Ds/FlareSense-v2} and the open-source dataset is available at \url{https://huggingface.co/datasets/i4ds/ecallisto_radio_sunburst}.

By releasing our dataset, model, and code as open source, we aim to promote reproducibility and accelerate collaborative advancements in the field of solar radio burst detection. Long-term support is planned through continued integration with the e-Callisto network and contributions from the community. We intend FlareSense to serve as a solid foundation for future work on solar radio burst detection and classification.

\section{Outlook}
Building on FlareSense's promising results, several areas for improvement and expansion emerge. Most urgently, real-time deployment across the e-Callisto network would enable near-instant detection of solar radio bursts, thereby aiding space-weather forecasting. Beyond single-instrument input, incorporating multi-instrument fusion, by combining data from additional networks (e.g., LOFAR \citep{van2013lofar} or MWA \citep{Tingay2013MWA}), could enhance detection reliability and refine burst characterization.

While FlareSense currently performs binary detection, classifying burst types (I-V) would further enrich its scientific value by illuminating the full spectrum of solar activity. In particular, fusing FlareSense outputs with localization cues from time-of-arrival pipelines such as BELLA \citep{Canizares_2024}, and exploiting the ever-growing, well-labelled e-Callisto archive, could supply the contextual information needed to train the network to discriminate reliably between burst morphologies and thus predict the specific burst type. For rare classes such as Type IV/V, few-shot approaches may help: Jiang et al. \citeyearpar{Jiang2026FewShot} recently applied few-shot object detection to Type III identification on STEREO/WAVES data; their detection code and dataset are not released, though they open-source their annotation tool.

In addition, handling noisy labels through refined annotations or semi-supervised learning techniques could boost performance, especially for faint or short-lived bursts. Building on model explainability, interactive interfaces that display attribution maps would also help observers validate detections and improve overall reliability.

By pursuing these directions, FlareSense can evolve into a robust, real-time platform for solar radio burst monitoring, advancing both scientific research and operational space weather services.

Furthermore, a more detailed investigation into how the burst-to-non-burst ratio during data generation and the tuning of class weights affects performance was deemed necessary. Generally, assigning a higher weight to burst samples usually improves recall, at the cost of precision.

Exploring methods to leverage the vast amount of unlabeled data provided by e-Callisto represents an exciting avenue for future study.

\section*{Acknowledgements}
We are particularly grateful to Arnold O. Benz for generously sharing his expertise on solar radio spectrograms, which provided the foundation for this study. We also extend our sincere thanks to Brandon Panos and Simon Felix for their valuable proofreading and thoughtful feedback on this manuscript.

\section*{Funding}
The e-Callisto network is a community-driven effort supported by funding from multiple nations. We especially acknowledge the support of the “Programa Giner de los Ríos 2024-2025” for Invited Researchers at Universidad de Alcalá (UAH), and the ESA Space Weather Network (SWESNET), which partially funded the data provision.

\section*{Data Availability}
The dataset used to produce FlareSense is available with an MIT license on Hugging Face at \url{https://huggingface.co/datasets/i4ds/ecallisto_radio_sunburst}. The demo is available at \url{https://flaresense.ch}.

\bibliographystyle{unsrtnat}
\bibliography{references}  

\appendix
\section{Stations and Instruments Used}
\label{appendix:stations_instruments_combined}
\begin{table}[H]
\centering
\caption{Stations and corresponding instruments used to generate and evaluate FlareSense data.}
\begin{tabular}{p{0.65\textwidth}p{0.35\textwidth}}
\hline
\textbf{Station} & \textbf{Instrument} \\
\hline
Cohoe Station (Alaska, USA) & ALASKA-COHOE\_63 \\
HAARP (High Frequency Active Auroral Research Program) (USA) & ALASKA-HAARP\_62 \\
CRAAG Station (Algeria) & ALGERIA-CRAAG\_59 \\
Almaty Observatory (Almaty, Kazakhstan) & ALMATY\_58 \\
University of Graz Radio Observatory (Austria) & AUSTRIA-UNIGRAZ\_01 \\
ASSA Station (Australia) & Australia-ASSA\_02 \& 62\\
Birr Castle Observatory (Ireland) & BIR\_01 \\
Alexandria Observatory (Egypt) & EGYPT-Alexandria\_02 \\
DLR Radio Station (Germany) & GERMANY-DLR\_63 \\
Glasgow Radio Observatory (United Kingdom) & GLASGOW\_01 \\
Humain Observatory (Belgium) & HUMAIN\_59 \\
Gauri Station (India) & INDIA-GAURI\_01 \\
Ooty Radio Telescope (India) & INDIA-OOTY\_02 \\
KASI Radio Station (South Korea) & KASI\_59 \\
MEXART (Mexican Array Radio Telescope) (Mexico) & MEXART\_59 \\
FCFM-UANL Radio Station (Mexico) & MEXICO-FCFM-UANL\_01 \\
LANCE-B Station (Mexico) & MEXICO-LANCE-B\_62 \\
UB Station (University of Ulaanbaatar) (Mongolia) & MONGOLIA-UB\_01 \\
Murchison Radio Observatory (Australia) & MRO\_59, MRO\_61 \\
Egersund Station (Norway) & NORWAY-EGERSUND\_01 \\
SSRT (Siberian Solar Radio Telescope) (Russia) & SSRT\_59 \\
Landschlacht Station (Switzerland) & SWISS-Landschlacht\_62 \\
Trieste Radio Observatory (Italy) & TRIEST\_57 \\
ERAU Radio Station (Embry-Riddle Aeronautical University) (USA) & USA-ARIZONA-ERAU\_01 \\
\hline
\end{tabular}
\end{table}

\section{Per-instrument metrics}
\label{appendix:per_instrument_metrics}

\begin{table}[H]
\caption{Per-instrument F1, precision, and recall (\%)}
\label{tab:all_instr_metrics_percent}
\centering
\begin{tabular}{lccc}
\hline\hline
\textbf{Instrument} & \textbf{F1 (\%)} & \textbf{Precision (\%)} & \textbf{Recall (\%)} \\
\hline
ALASKA-COHOE\_63 & 85.9 & 96.6 & 77.3 \\
ALASKA-HAARP\_62 & 88.4 & 97.4 & 80.9 \\
ALGERIA-CRAAG\_59 & 80 & 94.4 & 69.4 \\
ALMATY\_58 & 80.8 & 85.1 & 76.8 \\
AUSTRIA-UNIGRAZ\_01 & 81.9 & 89.3 & 75.6 \\
Australia-ASSA\_02 & 85.1 & 91.5 & 79.6 \\
Australia-ASSA\_62 & 80.7 & 95.5 & 69.8 \\
BIR\_01 & 91.2 & 92 & 90.5 \\
EGYPT-Alexandria\_02 & 86.3 & 91.5 & 81.7 \\
GERMANY-DLR\_63 & 77.6 & 97.2 & 64.5 \\
GLASGOW\_01 & 73 & 99.5 & 57.6 \\
HUMAIN\_59 & 89.6 & 92.5 & 86.8 \\
INDIA-GAURI\_01 & 67.9 & 94.8 & 52.9 \\
INDIA-OOTY\_02 & 76.7 & 92.9 & 65.3 \\
KASI\_59 & 82.6 & 100 & 70.4 \\
MEXART\_59 & 78.3 & 88.2 & 70.3 \\
MEXICO-FCFM-UANL\_01 & 90.9 & 100 & 83.3 \\
MEXICO-LANCE-B\_62 & 70.8 & 100 & 54.8 \\
MONGOLIA-UB\_01 & 93.1 & 97.9 & 88.7 \\
MRO\_59 & 61.9 & 68.4 & 56.5 \\
MRO\_61 & 74.9 & 88.5 & 64.9 \\
NORWAY-EGERSUND\_01 & 72.6 & 96.9 & 58 \\
SSRT\_59 & 75.4 & 91.1 & 64.3 \\
SWISS-Landschlacht\_62 & 77.4 & 90.2 & 67.8 \\
TRIEST\_57 & 92.4 & 94.4 & 90.5 \\
USA-ARIZONA-ERAU\_01 & 78.6 & 96.4 & 66.4 \\
\hline
\end{tabular}
\end{table}

\section{Data Generation Pipelines}

\begin{algorithm}[H]
\caption{Generate Burst Spectrograms}
\label{alg:generate_burst_spectro}
\begin{algorithmic}[1]
\Require Instrument list $\mathcal{I}$; burst catalog $\mathcal{B}$; offset range $[0,10]$ minutes; minimum number of frequency channels $f_{\min}{=}150$; duration threshold $d_{\min}{=}14$ minutes
\Ensure Set $\mathcal{D}_{\text{burst}}$ of labeled burst spectrograms
\For{each instrument $x \in \mathcal{I}$}
    \State $L_x \gets \textsc{FilterByInstrument}(\mathcal{B}, x)$
    \For{each entry $e \in L_x$}
        \State $t_0 \gets e.\text{datetime}$
        \State $\delta \gets \textsc{RandomOffset}(0, 10\ \text{min})$
        \State $\tau \gets t_0 - \delta$ \Comment{small negative offset to capture pre-onset context}
        \State $\text{data} \gets \textsc{LoadData}(x,\ \tau)$
        \If{$\textsc{FreqCount}(\text{data}) \ge f_{\min}$ \textbf{and} $\textsc{Duration}(\text{data}) \ge d_{\min}$}
            \State $\textsc{SaveSpectrogram}(\text{data};\ \text{label}{=}\texttt{"burst"})$
        \EndIf
    \EndFor
\EndFor
\end{algorithmic}
\end{algorithm}

\begin{algorithm}[H]
\caption{Generate Non-Burst Spectrograms}
\label{alg:generate_non_burst_spectro}
\begin{algorithmic}[1]
\Require Instrument list $\mathcal{I}$; burst catalog $\mathcal{B}$, each entry giving a burst time range and the stations that observed it; date range $[t_{\mathrm{start}}, t_{\mathrm{end}}]$; window length $\Delta{=}15$ minutes; multiplier $k{=}10$; minimum number of frequency channels $f_{\min}{=}150$; duration threshold $d_{\min}{=}14$ minutes
\Ensure Set $\mathcal{D}_{\text{non}}$ of labeled non-burst spectrograms
\State $\mathcal{T} \gets \textsc{BurstIntervals}(\mathcal{B})$ \Comment{time ranges of all catalog bursts, from \emph{any} station}
\For{each instrument $x \in \mathcal{I}$}
    \State $n_x \gets |\textsc{FilterByInstrument}(\mathcal{B}, x)|$ \Comment{number of catalog bursts for instrument $x$}
    \For{$i = 1$ \textbf{to} $k \cdot n_x$}
        \State $\tau \gets \textsc{RandomDate}(t_{\mathrm{start}}, t_{\mathrm{end}})$
        \If{$[\tau,\ \tau + \Delta]$ overlaps any interval in $\mathcal{T}$}
            \State \textbf{continue} \Comment{a burst was reported somewhere at this time; discard the window}
        \EndIf
        \State $\text{data} \gets \textsc{LoadData}(x,\ \tau)$
        \If{$\textsc{Available}(\text{data})$ \textbf{and} $\textsc{FreqCount}(\text{data}) \ge f_{\min}$ \textbf{and} $\textsc{Duration}(\text{data}) \ge d_{\min}$}
            \State $\textsc{SaveSpectrogram}(\text{data};\ \text{label}{=}\texttt{"non-burst"})$
        \EndIf
    \EndFor
\EndFor
\end{algorithmic}
\end{algorithm}

\section{Explainable AI Examples}
\label{appendix:ex_ai_examples}
\noindent\textbf{Attribution examples.} The XAI examples were generated before temperature scaling and therefore use the uncalibrated probability threshold of 0.5. The matched-precision evaluation in Section~\ref{sec:evaluation}, by contrast, uses temperature-scaled probabilities and the calibrated threshold of 0.426. Captions emphasize what the attribution maps reveal about the image regions associated with each prediction.

\begin{figure}[H]
    \centering
    \includegraphics[width=1\textwidth]{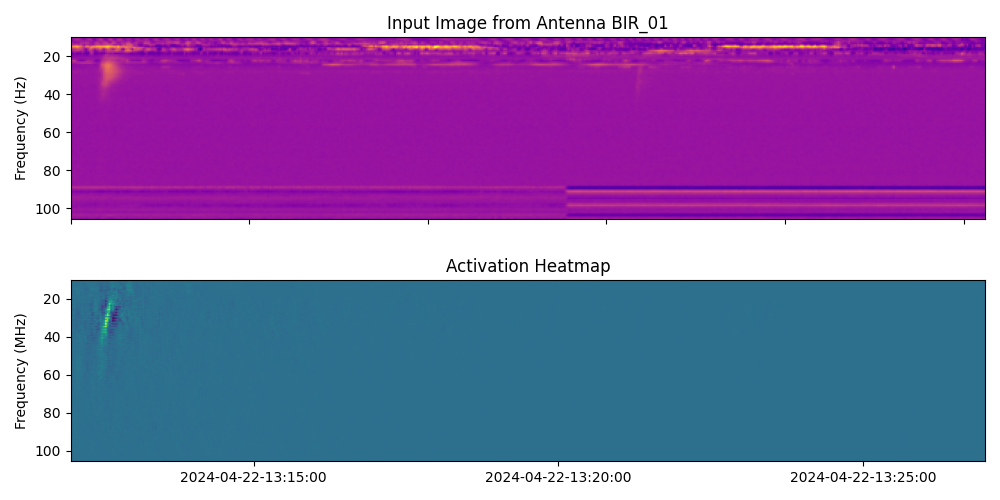}
    \caption{\small True positive - \texttt{BIR\_01}. Attribution concentrates on the burst ridge and suppresses background structure.}
    \label{fig:xai_1}
\end{figure}

\begin{figure}[H]
    \centering
    \includegraphics[width=1\textwidth]{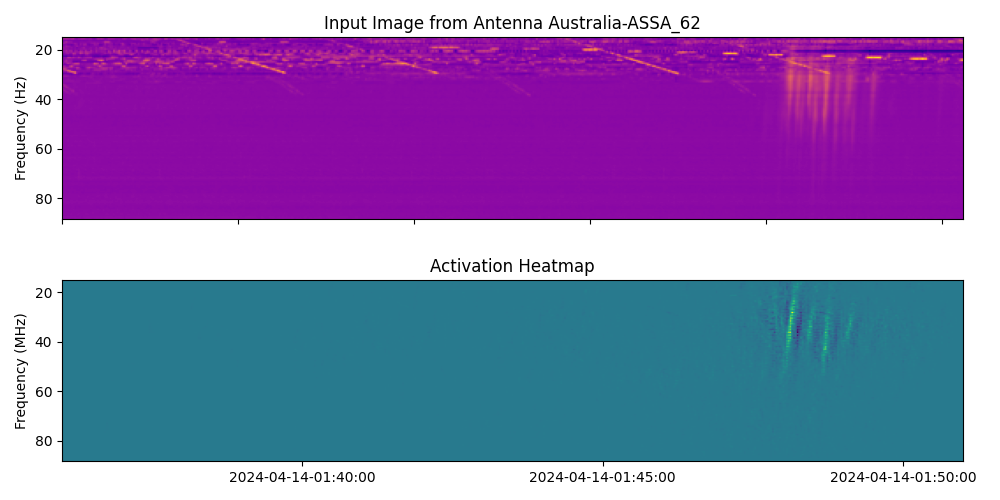}
    \caption{\small True positive - \texttt{Australia-ASSA\_62}. Attribution follows the salient emission trace, showing consistent behavior across instruments.}
    \label{fig:xai_2}
\end{figure}

\begin{figure}[H]
    \centering
    \includegraphics[width=1\textwidth]{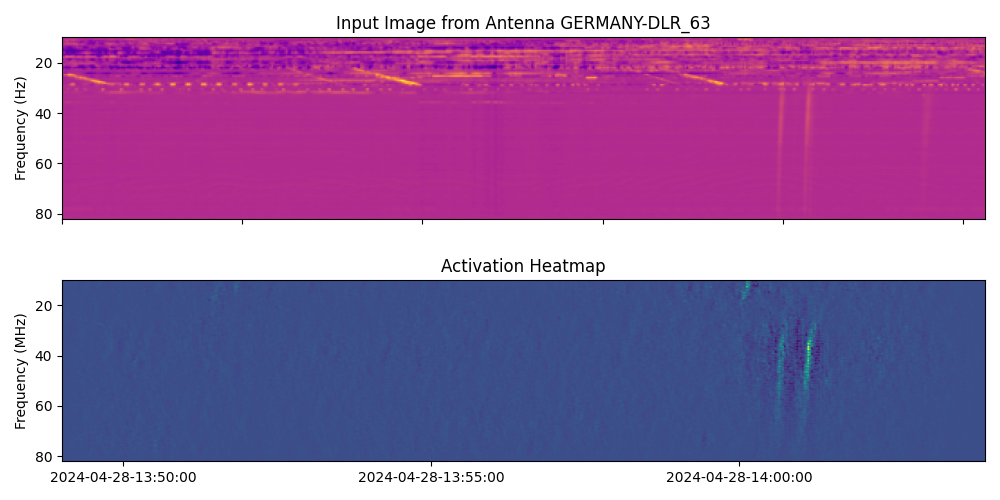}
    \caption{\small True positive - \textit{Germany-DLR\_63}. Attribution localizes along the drift-like burst pattern.}
    \label{fig:xai_3}
\end{figure}

\begin{figure}[H]
    \centering
    \includegraphics[width=1\textwidth]{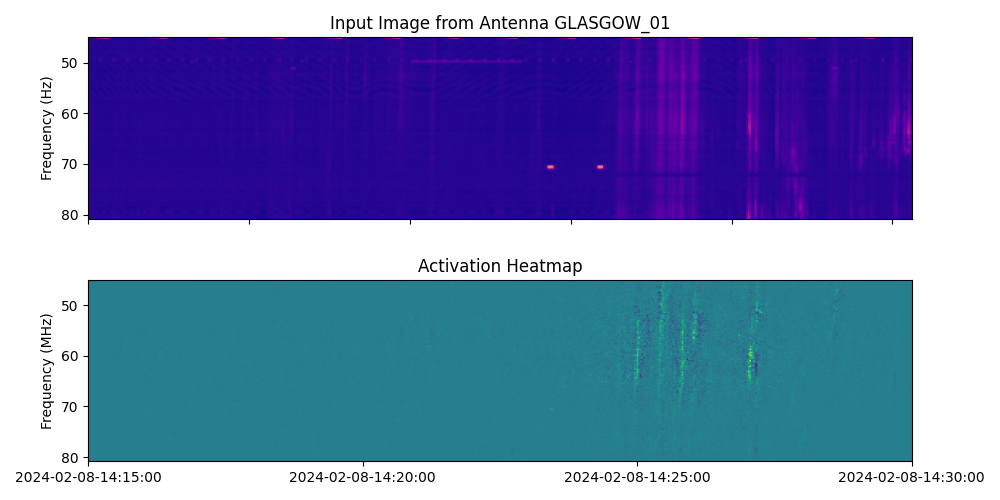}
    \caption{\small True positive - \textit{Glasgow\_01}.}
    \label{fig:xai_4}
\end{figure}

\begin{figure}[H]
    \centering
    \includegraphics[width=1\textwidth]{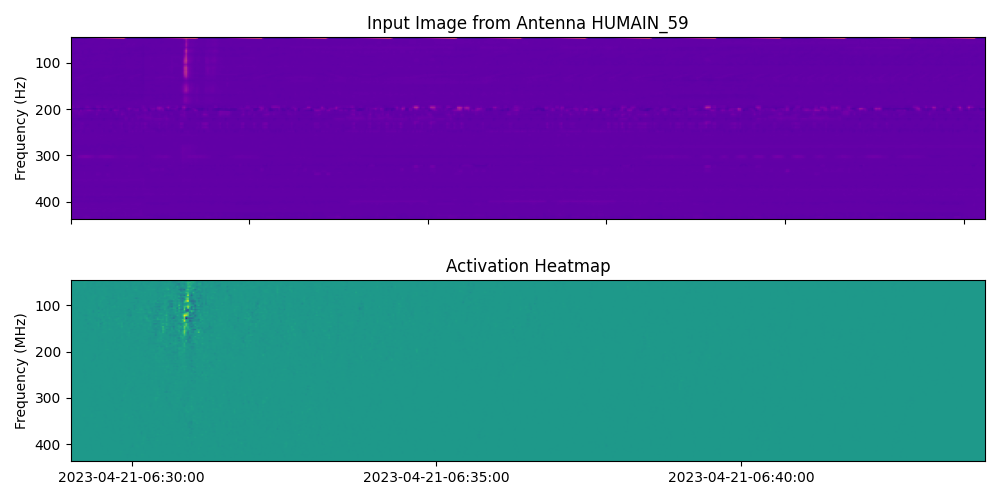}
    \caption{\small True positive - \textit{Humain\_59}.}
    \label{fig:xai_5}
\end{figure}

\begin{figure}[H]
    \centering
    \includegraphics[width=1\textwidth]{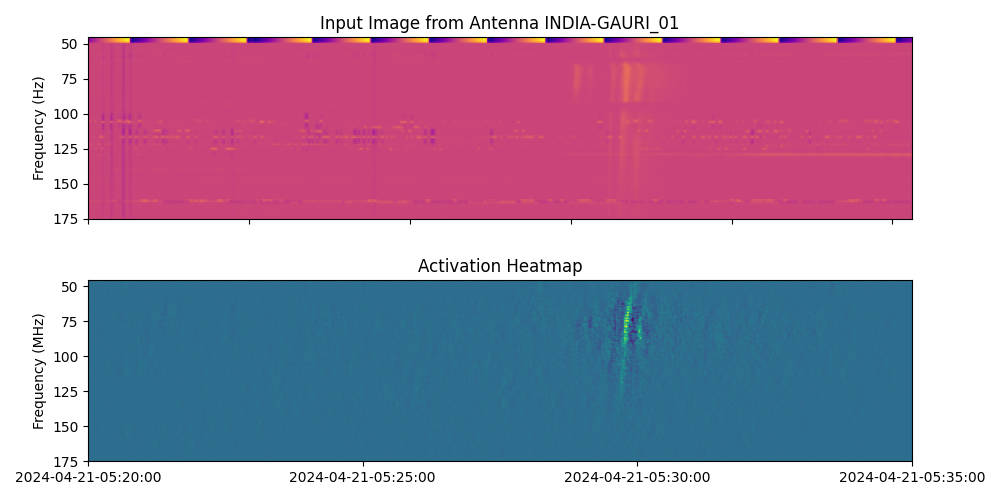}
    \caption{\small False positive - \textit{India-Gauri\_01}. This is another example for an annotation miss (see Section \ref{sec:model_robust}), so in effect, this case is another true positive.}
    \label{fig:xai_6}
\end{figure}

\begin{figure}[H]
    \centering
    \includegraphics[width=1\textwidth]{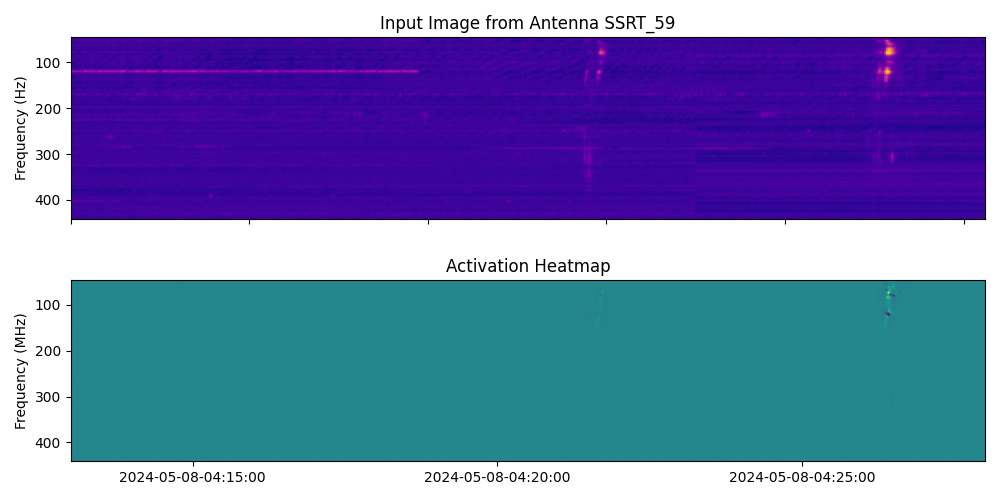}
    \caption{\small True positive - \textit{SSRT\_59}. Strong attribution along the emission band indicates sensitivity to real events.}
    \label{fig:xai_7}
\end{figure}
\section{Probability Calibration}
\label{app:prob_calib}
Neural-network confidence scores are not necessarily calibrated probabilities \citep{guo2017calibrationmodernneuralnetworks}. Additionally, label smoothing \citep{szegedy2016rethinking} makes the distribution of the output probabilities less sharp. We therefore fitted a single temperature parameter by minimizing the negative log likelihood of the predictions on the training set. The resulting temperature was 0.4974. Because this value is below one, temperature scaling sharpens the probability distribution, counteracting part of the smoothing introduced during training.
\begin{figure}[H]
    \centering
    \includegraphics[width=0.6\textwidth]{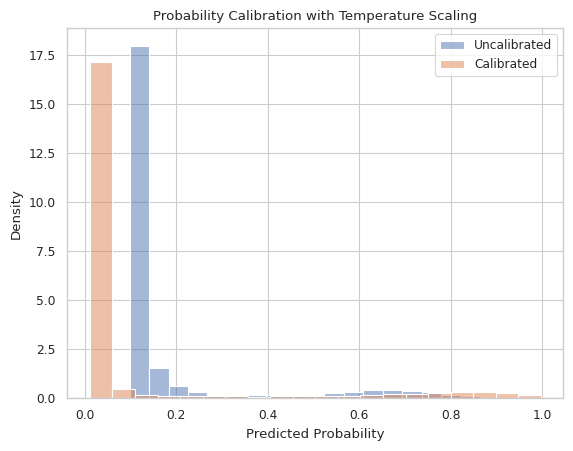}
    \caption{Distributions of model probabilities before and after temperature scaling. The fitted temperature of 0.4974 sharpens the probabilities relative to the uncalibrated outputs.}
    \label{fig:prob_calib}
\end{figure}

\end{document}